\documentclass[twocolumn,aps,showpacs,amsmath,superscriptaddress]{revtex4-1}

\usepackage{bm}
\usepackage{amsmath}
\usepackage{amssymb}
\usepackage[usenames,dvipsnames]{color}
\usepackage{graphicx}
\usepackage{dcolumn}
\usepackage{simplewick}
\usepackage{hyperref}
\hypersetup{
  colorlinks=true,
  citecolor=blue,
  linkcolor=blue,
  urlcolor=blue}

\begin{document}

\title{Electric dipole polarizability of alkaline-Earth-metal atoms from
       perturbed relativistic coupled-cluster theory with triples}
\author{S. Chattopadhyay}
\affiliation{Physical Research Laboratory,
             Ahmedabad - 380009, Gujarat, 
             India}
\author{B. K. Mani}
\affiliation{Department of Physics, University of South Florida, Tampa,
             Florida 33620, USA}
\author{D. Angom}
\affiliation{Physical Research Laboratory,
             Ahmedabad - 380009, Gujarat,
             India}

\begin{abstract}
  The perturbed relativistic coupled-cluster (PRCC) theory is applied to
calculate the electric dipole polarizabilities of alkaline Earth metal atoms. 
The Dirac-Coulomb-Breit atomic Hamiltonian is used and we include the triple 
excitations in the relativistic coupled-cluster (RCC) theory. The theoretical 
issues related to the triple excitation cluster operators are described in 
detail and we also provide details on the computational implementation. The
PRCC theory results are in good agreement with the experimental and previous
theoretical results. We, then, highlight the importance of considering
the Breit interaction for alkaline Earth metal atoms.
\end{abstract}



\maketitle


\section{Introduction}

 Static electric dipole polarizability, $\alpha$ of the neutral alkaline 
Earth metal atoms \cite{bonin-97} is important for many applications in the 
ongoing and future experiments. To name a few, the parity and time reversal 
violation in atoms \cite{dzuba-00,ginges-04}, optical atomic clocks 
\cite{ludlow-08,wilpers-07} and  condensates of dilute atomic gases
\cite{escobar-09,stellmer-09,kraft-09} are of current interest. 
In the past different many-body methods have been used to calculate $\alpha$ 
precisely. Some of the theoretical methods used to calculate $\alpha$ are
described in a recent review by Mitroy {\em et. al} \cite{mitroy-10}. Another 
important and excellent reference for the ground state $\alpha$ of neutral 
atoms is the table of experimental data and theoretical results prepared by 
Schwerdtfeger \cite{schwerdtfeger-06}.  In this paper 
we employ the perturbed relativistic coupled-cluster (PRCC) theory 
\cite{chattopadhyay-12a, chattopadhyay-12b} to calculate the $\alpha$ of 
alkaline Earth metal atoms. The theoretical details of PRCC theory are
discussed in our previous works
\cite{chattopadhyay-12a, chattopadhyay-12b, chattopadhyay-13a, 
chattopadhyay-13b}. A related method used for calculating electric dipole 
polarizabilities is to consider only the $z$-component of the dipole operator 
and define a set of perturbed cluster operators \cite{sahoo-08}.
In PRCC theory we introduce a new set of cluster operators
along with the original RCC operators. For convenient description we refer to 
the latter as the unperturbed cluster operators. The technique is general and 
suitable to incorporate multiple perturbations for structure and properties 
calculations of many electron atoms and ions. So, in the PRCC theory, the 
cluster operators can be scalar or tensor operators of any rank and this is
an important feature of the PRCC theory we have developed.  

 Among the different many body techniques the coupled-cluster theory
(CCT) based methods are efficient and powerful. A recent review 
\cite{bartlett-07} on the CCT gives an insight about the theory and related
CCT based techniques. The CCT has been widely used for atomic 
\cite{mani-09,latha-09, nataraj-08,pal-07,geetha-01}, 
molecular \cite{isaev-04}, nuclear \cite{hagen-08} and condensed matter 
physics \cite{bishop-09} calculations. In the present work, we introduce the 
triple excitations in the RCC theory to go beyond coupled-cluster single and 
double (CCSD) approximation. The inclusion of the triple excitations 
incorporate some of the many-body effects equivalent to the diagrams which 
begin to contribute from the second order in many-body perturbation theory 
(MBPT). The triple excitations at the RCC theory shall pave the way for high 
precision results for atomic structure calculations. It must be mentioned that, 
a previous work have considered the triple excitation cluster amplitudes
in RCC calculations \cite{derevianko-08}. In this work we, however, introduce
a different but equivalent representation of the triple excitation cluster
operator. So, in the present work, the unperturbed clusters are calculated
with the relativistic coupled-cluster single, double and triple (RCCSDT) 
excitation approximation. We, however, use the PRCC theory with single and 
double excitation approximation. In our future work we shall include the 
effects of triple excitations in the PRCC theory.
 
 The paper is organized as follows. In the Sec. II. A, we briefly introduce the 
RCC and PRCC theories with the Dirac-Coulomb-Breit Hamiltonian. In Sec. II. B, 
we introduce the triple excitations $(T_3^{(0)})$ in the RCC theory and derive
the RCCSDT amplitude equations. In the next part we introduce the 
angular momentum diagrams corresponding to $T_3^{(0)}$ operator and 
evaluation of $T_3^{(0)}$ diagrams in the RCC theory. In
Sec. III we briefly discuss about the $\alpha$ calculations
in the framework of PRCC theory. In the subsequent sections we describe the 
calculational details and the computational issues related to $T_3^{(0)}$ 
calculations. We, then, present the results and discussions for neutral
alkaline Earth metal atoms and end with conclusions. All the results 
presented in this work and related calculations are in atomic units
( $\hbar=m_e=e=1/4\pi\epsilon_0=1$). In this system of units the velocity of
light is $\alpha ^{-1}$, the inverse of fine structure constant. For which we
use the value of $\alpha ^{-1} = 137.035\;999\;074$ \cite{mohr-12}.


\section{Theoretical methods}

For the high-$Z$ atoms and ions, the Dirac-Coulomb-Breit Hamiltonian, denoted 
by $H^{\rm DCB}$, is an appropriate choice to include the relativistic effects. 
However, there are complications associated with the negative energy 
continuum states of $H^{\rm DCB}$. These lead to variational collapse and 
{\em continuum dissolution} \cite{brown-51}. A formal approach to avoid these 
complications is to use the no-virtual-pair approximation. In this 
approximation, for a neutral atom of $N$ electrons \cite{sucher-80}
\begin{eqnarray}
   H^{\rm DCB} & = & \Lambda_{++}\sum_{i=1}^N \left [c\bm{\alpha}_i \cdot 
        \mathbf{p}_i + (\beta_i -1)c^2 - V_{N}(r_i) \right ] 
                       \nonumber \\
    & & + \sum_{i<j}\left [ \frac{1}{r_{ij}}  + g^{\rm B}(r_{ij}) \right ]
        \Lambda_{++},
\end{eqnarray}
where $\bm{\alpha}$ and $\beta$ are the Dirac matrices, $\Lambda_{++}$ is an 
operator which projects to the positive energy solutions and $V_{N}(r_{i})$ is 
the nuclear potential. Sandwiching the Hamiltonian with $\Lambda_{++}$ ensures 
that the effects of the negative energy continuum  states are neglected in the 
calculations. Another approach, which is better suited for numerical
computations, is to use the kinetically balanced finite basis sets
\cite{stanton-84,mohanty-90,grant-06,grant-10}. We use this method in the
present work to generate the orbital basis sets.
The last two terms, $1/r_{ij} $ and $g^{\rm B}(r_{ij})$  are the 
Coulomb and Breit interactions, respectively.  The latter, Breit interaction, 
represents the inter-electron magnetic interactions and is given by
\begin{equation}
  g^{\rm B}(r_{12})= -\frac{1}{2r_{12}} \left [ \bm{\alpha}_1\cdot\bm{\alpha}_2
               + \frac{(\bm{\alpha_1}\cdot \mathbf{r}_{12})
               (\bm{\alpha_2}\cdot\mathbf{r}_{12})}{r_{12}^2}\right].
\end{equation}
The Hamiltonian satisfies the eigen-value equation
\begin{equation}
   H^{\rm DCB}|\Psi_{i}\rangle = E_{i}|\Psi_{i}\rangle , 
\end{equation}
where, $|\Psi_{i}\rangle$ is the exact atomic state and $E_i$ is the energy 
of the atomic state. In the presence of external electromagnetic fields, the
Hamiltonian is modified with the addition of interaction terms. For external 
static electric field, the interaction is 
$H_{\rm int}=-\mathbf{d}\cdot\mathbf{E}_{\rm ext} $, 
where $\mathbf{d}$ and $\mathbf{E}_{\rm ext}$ are the induced electric dipole 
moment of the atom and external electric field, respectively. In the remaining
part of this section we give a brief description of RCC theory, which we use
to compute atomic state $|\Psi\rangle $ and PRCC to account for the 
effects $H_{\rm int}$ in the atomic state.


\subsection{RCC and PRCC theories}

 In RCC theory we define the  ground state atomic wavefunction of a 
closed-shell atom as 
\begin{equation}
|\Psi_0\rangle = e^{ T^{(0)}}|\Phi_0\rangle ,
\end{equation}
where $|\Phi_0\rangle$ is the reference state wave-function and $T^{(0)}$ is 
the unperturbed cluster operator. To account for the correction to 
the wavefunction arising from $H_{\rm int} $, we define the perturbed ground 
state as
\begin{equation}
 |\tilde{\Psi}_0\rangle = e^{T^{(0)} + \lambda \mathbf{T}^{(1)}\cdot\mathbf{E}} 
 |\Phi_0\rangle = e^{T^{(0)}}\left [ 1 + \lambda \mathbf{T^{(1)}\cdot 
 \mathbf{E}} \right ] |\Phi_0\rangle , \;\;\;\;
 \label{psi_tilde}
\end{equation}
where $\mathbf{T}^{(1)}$ are the PRCC operators 
\cite{chattopadhyay-12a,chattopadhyay-12b}. For an $N$-electron 
closed-shell atom $T^{(0)} = \sum_{i=1}^N T_{i}^{(0)}$ and 
${\mathbf T}^{(1)} = \sum_{i=1}^N {\mathbf T}_{i}^{(1)}$, where $i$ is the 
order of excitation.  In the coupled-cluster single and double (CCSD) 
excitation approximation \cite{purvis-82},
\begin{subequations}
\begin{eqnarray}
T^{(0)} &=& T_{1}^{(0)} + T_{2}^{(0)}, \\ 
{\mathbf T}^{(1)} &=& {\mathbf T}_{1}^{(1)} + {\mathbf T}_{2}^{(1)}.
\end{eqnarray}
\end{subequations}
The CCSD is a good starting point for structure and properties calculations 
of closed-shell atoms and ions. In the second quantized representation
\begin{subequations}
\begin{eqnarray}
  T_1^{(0)} &= &\sum_{a,p} t_a^p {{a}_p^\dagger} a_a , \\
  T_2^{(0)} &= &\frac{1}{(2!)^2}\sum_{a,b,p,q}
                t_{ab}^{pq} {{a}_p^\dagger}{{a}_q^\dagger}a_b a_a , \\
  \mathbf{T}_1^{(1)} & = & \sum_{a,p} \tau_a^p \mathbf{C}_1 (\hat{r})
                       a_{p}^{\dagger}a_{a},
                            \\
  \mathbf{T}_2^{(1)} & = & \frac{1}{(2!)^{2}}\sum_{a,b,p,q} 
                   \sum_{l,k} \tau_{ab}^{pq}(l,k) 
                   \{ \mathbf{C}_l(\hat{r}_1) \mathbf{C}_k(\hat{r}_2)\}^{1}
                   a_{p}^{\dagger}a_{q}^{\dagger}a_{b}a_{a}, \;\;\;\;\;\;\;\;
\end{eqnarray}
\end{subequations}
where $t_{\ldots}^{\ldots}$ and $\tau_{\ldots}^{\ldots}$ are the cluster 
amplitudes, $a_i^{\dagger}$ ($a_i$) are single particle creation (annihilation)
operators and $abc\ldots$ ($pqr\ldots$) represent core (virtual) single 
particle states or orbitals. To represent $\mathbf{T}_1^{(1)}$, a rank one 
operator, we have used the $\mathbf{C}$-tensor of similar rank 
$\mathbf{C}_1(\hat r)$. Coming to $\mathbf{T}_2^{(1)}$, to represent it two 
$\mathbf{C}$-tensor operators of rank $l$ and $k$ are coupled to a rank one 
tensor operator. In addition, the PRCC clusters are constrained by 
other selection rules arising from parity and triangular conditions, 
these are described in our previous work \cite{chattopadhyay-12b}.
\begin{figure}
   \includegraphics[width=5.5cm]{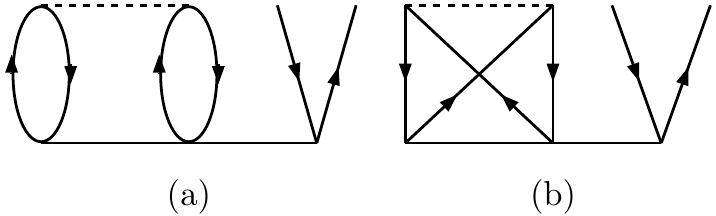}
   \caption{Diagrams of $T_1^{(0)}$ cluster operator arising from the triple
            excitation cluster operator, $T_3^{(0)}$.
           }
   \label{single_triple}
\end{figure}

\begin{figure}
   \includegraphics[width=8.5cm]{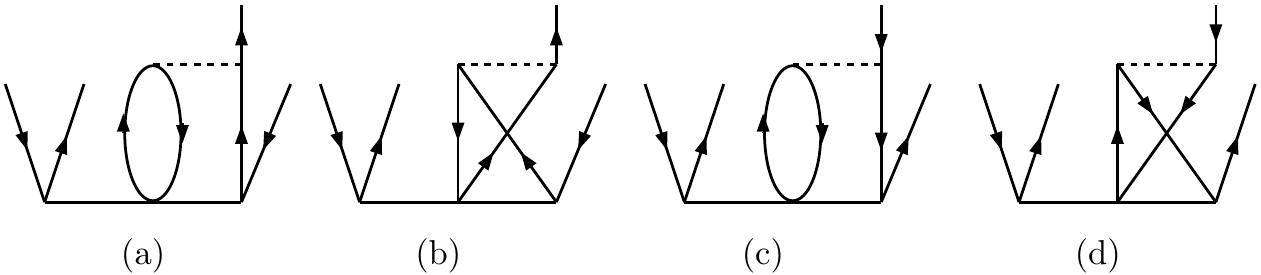}
   \caption{Diagrams of $T_2^{(0)}$ cluster operator arising from the triple
            excitation cluster operator, $T_3^{(0)}$.
           }
   \label{double_triple}
\end{figure}


\subsection{CCSDT approximation}

\begin{figure}[h]
   \includegraphics[width=9.0cm]{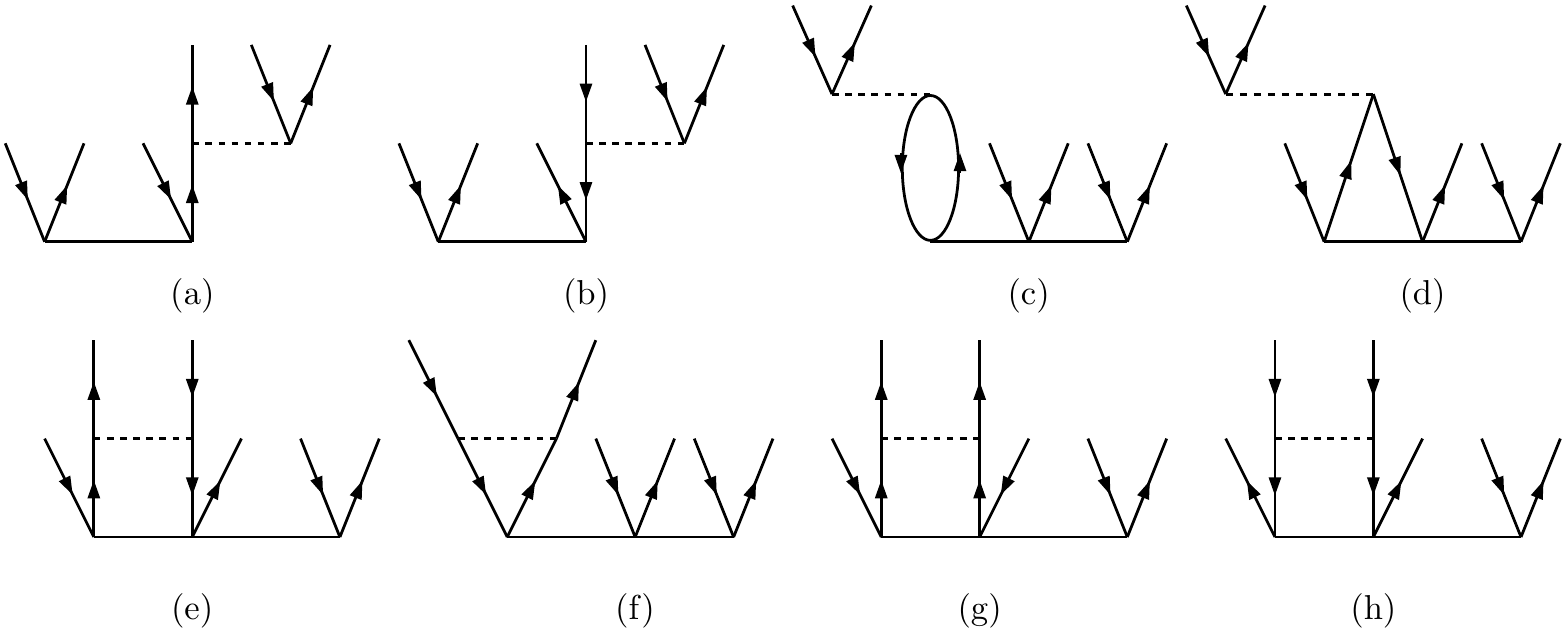}
   \caption{Diagrams of $T_3^{(0)}$ arising from $T_3^{(0)}$ in the 
            linearized RCCSDT theory. 
           }
   \label{t3_diag}
\end{figure}

 The RCCSD approximation encompasses a major part of the electron correlation
effects. It is, however, pertinent to incorporate triple excitation cluster
operator, $T_3^{(0)}$, to obtain higher precision. With $T_3^{(0)}$, the 
theory is referred to as the relativistic coupled-cluster single, double
and triple (RCCSDT) excitation approximation, then
$T^{(0)}= T_1^{(0)} + T_2^{(0)} + T_3^{(0)} $. In the second quantized 
notations, we may write the triple excitation cluster operator as
\begin{equation}
  T_3^{(0)} = \frac{1}{(3!)^2}\sum_{\substack { a,b,c \\ p,q,r}}
              t_{abc}^{pqr} {{a}_p^\dagger}{{a}_q^ \dagger}{{a}_r^ \dagger}
              a_ca_b a_a ,
\end{equation}
and the cluster operators are the solutions of the equation
\begin{equation}
  \langle\Phi^{pqr}_{abc}|\bar H_{\rm N}^{\rm DCB}|\Phi_0\rangle = 0.
  \label{cct3}
\end{equation}
Where, $H_{\rm N}^{\rm DCB}$ is the normal ordered Dirac-Coulomb-Breit 
Hamiltonian and 
$\bar{H}_{\rm N}^{\rm DCB}=e^{-T^{(0)}}H_{\rm N}^{\rm DCB}e^{T^{(0)}} $ is the
similarity transformed or dressed Hamiltonian. Here after, for compact 
notation we use $\bar{H}_{\rm N}$ to represent $\bar{H}_{\rm N}^{\rm DCB}$. 
Following Wick's theorem and structure of $H_{\rm N}$, in general
\begin{eqnarray}
  \bar{H}_{\rm N}=&&H_{\rm N}+\{\contraction[0.4ex]{}{H}{_{\rm N}}{T}
   H_{\rm N}T^{(0)}\} +
  \frac{1}{2!}\{\contraction[0.4ex]{}{H}{_{\rm N}}{T}
  \contraction[0.7ex]{}{H}{_{\rm N}T^{(0)}}{T} H_{\rm N}T^{(0)}T^{(0)}\} +
  \nonumber \\
  &&\frac{1}{3!}\{\contraction[0.4ex]{}{H}{_{\rm N}}{T}
  \contraction[0.7ex]{}{H}{_{\rm N}T^{(0)}}{T}
  \contraction[1.0ex]{}{H}{_{\rm N}T^{(0)}T^{(0)}}{T}
  H_{\rm N}T^{(0)}T^{(0)}T^{(0)}\}  
  + \frac{1}{4!}\{\contraction[0.4ex]{}{H}{_{\rm N}}{T}
  \contraction[0.7ex]{}{H}{_{\rm N}T^{(0)}}{T}
  \contraction[1.0ex]{}{H}{_{\rm N}T^{(0)}T^{(0)}}{T}
  \contraction[1.3ex]{}{H}{_{\rm N}T^{(0)}T^{(0)}T^{(0)}}{T}
  H_{\rm N}T^{(0)}T^{(0)}T^{(0)}T^{(0)}\}, \nonumber
  \label{Hnbarcont}
\end{eqnarray}
where $\contraction[0.4ex]{}{A}{\ldots}{B}A\ldots B$ denotes contraction 
between the two operators $A$ and $B$, and $\{\cdots\}$ represent normal
ordering of the operator. A detailed description of the CCSDT cluster 
equations, in the context of non-relativistic systems, is given the 
in the recent book of Shavitt and Bartlett \cite{shavitt-09}. The equations
can be modified to the relativistic case. Although, we solve
the full RCCSDT equations, for a more compact description but to provide all 
the key details, here we give the description of linearised RCCSDT (LRCCSDT).
In this approximation we retain terms in the $T^{(0)}$ equations which are 
zeroth and first order in $T^{(0)}$. It is an approximation which is 
relatively simple but incorporates all the important many-body effects. 
In addition, solving the LRCCSDT equations is not computationally intensive, 
for this reason, we solve these equations first and use the results as 
starting values to solve the full or nonlinear RCCSDT equations. In the 
LRCCSDT the cluster amplitude equations are then
\begin{subequations}
  \begin{equation}
  \langle\Phi^p_a|H_{\rm N} + \{\contraction[0.4ex]{}{H}{_{\rm N}}{T}H_{\rm N} 
   T^{(0)}_1\} + \{\contraction[0.4ex]{}{H}{_{\rm N}}{T}H_{\rm N} T^{(0)}_2\}
   + \{\contraction[0.4ex]{}{H}{_{\rm N}}{T}H_{\rm N} T^{(0)}_3\}  
   |\Phi_0\rangle = 0, 
   \label{ccsdt_sing} 
  \end{equation}
  \begin{equation}
  \langle\Phi^{pq}_{ab}|H_{\rm N} + \{\contraction[0.4ex]{}{H}{_{\rm N}}{T}
   H_{\rm N} T^{(0)}_1\} + \{\contraction[0.4ex]{}{H}{_{\rm N}}{T}H_{\rm N} 
   T^{(0)}_2\} + \{\contraction[0.4ex]{}{H}{_{\rm N}}{T}H_{\rm N} T^{(0)}_3\}  
   |\Phi_0\rangle = 0, 
   \label{ccsdt_dbl} 
  \end{equation}
  \begin{equation}
  \langle\Phi^{pqr}_{abc}|\{\contraction[0.4ex]{}{H}{_{\rm N}}{T}H_{\rm N} 
   T^{(0)}_2\} + \{\contraction[0.4ex]{}{H}{_{\rm N}}{T}H_{\rm N} T^{(0)}_3\}  
   |\Phi_0\rangle = 0. 
   \label{ccsdt_tripl} 
  \end{equation}
\end{subequations}
Here, an important observation is the absence of $H_{\rm N}$ 
and $\{\contraction[0.4ex]{}{H}{_{\rm N}}{T} H_{\rm N} T^{(0)}_1\}$ in the
$T_3^{(0)}$ equation. The reason is, as $H_{\rm N}$ is a two-body interaction 
Hamiltonian, at first order it does not induce triple excitations by itself or
after contraction with $T_1^{(0)}$. 
  In the LRCCSDT,  $ T_1^{(0)}$ and $T_2^{(0)}$ have, respectively, 
two and four diagrams arising from from $T_3^{(0)}$. The latter are shown in 
Fig. \ref{double_triple} and these originate from two types of residual 
two-body interactions, namely $g^{ra}_{pq}$ and $g^{bc}_{ap}$, where 
$g_{ij}^{kl} = \langle kl|1/r_{12} + g^{B}(r_{12})|ij\rangle$. In Fig. 
\ref{double_triple}, the diagrams (a-b) and (c-d) arise from 
$g^{ra}_{pq}$ and $g^{bc}_{ap} $ respectively. Similarly, the diagrams of 
$T_3^{(0)}$ arising from $T_3^{(0)}$ in the linearized RCCSDT are shown in 
Fig. \ref{t3_diag}. There are eight diagrams and these arise from six types 
of residual Coulomb interactions $g^{qr}_{pa}$, $g^{cp}_{ab}$, $g^{pb}_{aq}$, 
$g^{qb}_{pa} $, $g^{rs}_{pq} $ and $g^{cd}_{ab}$.The contribution from 
$T_3^{(0)}$ to the $T_1^{(0)}$ cluster equation, Eq. (\ref{ccsdt_sing}), may 
be written in the algebraic form as 
\begin{equation}
   \langle \contraction[0.4ex]{}{H}{_{\rm N}}{T}H_{\rm N} 
    T^{(0)}_3 \rangle_{a}^{p} = \sum_{bcqr}(g^{bc}_{qr} - g^{bc}_{rq})
                                t_{abc}^{pqr}, \nonumber    
\end{equation}
where $\langle\cdots\rangle _a^p$ represents 
$\langle\Phi_a^p|\cdots|\Phi_0\rangle $. Similarly, the $T_3^{(0)}$ 
contribution to the $T_2^{(0)}$ cluster operator equation, fourth term in 
Eq. (\ref{ccsdt_dbl}), is
\begin{equation}
   \langle \contraction[0.4ex]{}{H}{_{\rm N}}{T}H_{\rm N} 
    T^{(0)}_3 \rangle_{ab}^{pq} = \sum_{rcs}(g^{rs}_{cq} - g^{sr}_{cq})
         t_{acb}^{prs} + \frac{1}{2} \sum_{rcd}
         (g^{rb}_{cd} - g^{rb}_{dc})t_{acd}^{prq} . 
    \nonumber 
\end{equation}
With these definitions of the terms arising from $T_3^{(0)}$, the 
$T_1^{(0)}$ cluster amplitude equation, Eq. (\ref{ccsdt_sing}), in algebraic 
form is
\begin{eqnarray}
    & & \sum_{bq} g^{bp}_{aq}t_b^q + \frac{1}{2}\sum_{bcq}g^{bc}_{qa}
    (t_{bc}^{qp} - t_{bc}^{pq}) 
    + \sum_{bqr}g^{bp}_{qr}(t_{ba}^{qr} - t_{ab}^{qr}) 
    \nonumber \\  
    & & + \frac{1}{2}\sum_{bcqr} (g^{bc}_{qr} - g^{bc}_{rq})t_{abc}^{pqr} 
    + \left (\varepsilon_p - \varepsilon_a \right ) t_a^p= 0,
\end{eqnarray}
where, $\varepsilon_i$ is the orbital energy of the $i$th orbital.
Similarly, the $T_2^{(0)}$ cluster amplitude equation, Eq. (\ref{ccsdt_dbl}),
can be written in algebraic form as
\begin{eqnarray}
   & & \sum_{r}g^{pq}_{ar}t_b^r - \sum_{c}g^{pc}_{ab}t_c^q 
   + \sum_{cd}g^{cd}_{ab}t_{cd}^{pq} + \sum_{rs}g^{pq}_{rs}t_{ab}^{rs} 
   - \sum_{cr} \bigg [ g^{cp}_{ar} t_{cb}^{rq}   
   \nonumber \\
   & & + g^{pc}_{rb}t_{ac}^{rq} + \frac{1}{2} g^{pc}_{ar}(t_{cb}^{rq} 
   - t_{bc}^{rq}) \bigg ] + \sum_{rcs}(g^{rs}_{cq} 
   - g^{sr}_{cq}) t_{acb}^{prs} + \frac{1}{2} \sum_{rcd}(g^{rb}_{cd}
   \nonumber \\
   & &   - g^{rb}_{dc})t_{acd}^{prq}  +  \left ( \begin{array}{c}
                                                   p \leftrightarrow q \\
                                                   a \leftrightarrow c
                                                 \end{array}\right )
    + \left (\varepsilon_p + \varepsilon_q 
   - \varepsilon_a - \varepsilon_b\right )t_{ab}^{pq}= 0,
\end{eqnarray}
where $i\leftrightarrow j$ represents permutation of the two indexes.
As evident from the $T_3^{(0)}$ cluster equation, Eq. (\ref{ccsdt_tripl}), 
the contributing terms are the contraction of $H_{\rm N}$ with $T_{2}^{(0)}$ 
and $T_{3}^{(0)}$. The Goldstone diagrams arising from these terms are shown 
in Fig. \ref{t3_diag}, where the diagrams in Fig. \ref{t3_diag}(a-b) arise
from $T_2^{(0)} $, and the diagrams in Fig. \ref{t3_diag}(c-h) arise from
$T_3^{(0)}$. Collecting all the diagrams, the equation of the $T_3^{(0)}$
cluster amplitude in algebraic form is
\begin{eqnarray}  
  && \sum_{s} g^{qr}_{sc}t_{ab}^{ps}   + \sum_{d} g^{dr}_{bc}t_{ad}^{pq}
    + \sum_{ds} \bigg [ g^{as}_{pd}\left ( t_{dbc}^{sqr} 
    + t_{bdc}^{sqr} \right )  + g^{sb}_{pd}t_{adc}^{sdr}
                 \nonumber \\
  &&+ g^{as}_{dp}t_{dbc}^{sqr} \bigg ] 
    + \sum_{st} g^{st}_{pq}t_{abc}^{str} + \sum_{de} g^{ab}_{de}t_{dec}^{pqr}
    + \left ( \begin{array}{c}
                 p \leftrightarrow q \leftrightarrow r\\
                 a \leftrightarrow b \leftrightarrow c
              \end{array}\right )
   \nonumber \\
   & & + \left (\varepsilon_p + \varepsilon_q + \varepsilon_r- \varepsilon_a 
    - \varepsilon_b - \varepsilon_c\right ) t_{abc}^{pqr}= 0.
\end{eqnarray}
With the inclusion of $T_3^{(0)}$ the RCC theory incorporates all the 
correlation effects  up to second order in the residual Coulomb interaction. 
That is, the theory encapsulates all the many-body perturbation theory (MBPT) 
diagrams \cite{lindgren-86} which are first and second order in the 
residual Coulomb interaction. In addition, as it is coupled cluster theory, 
it incorporates the connected single, double and triple excitations 
to all order. The leading order contribution to the uncertainty in the 
calculations arise from the quadruple excitations, which, in MBPT, first 
appear at the third order of perturbation.


\subsection{Representation of $T_3^{(0)} $ }

\begin{figure}[h]
   \includegraphics[width=5.0cm]{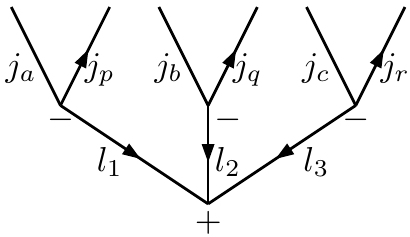}
   \caption{Angular momentum representation of $T_3^{(0)}$ cluster operator.
            The $-(+)$ sign indicates the angular momenta at the vertices are
            coupled in clock (anti-clock) wise sequence. An arrow, pointing
            away from the vertex, on a line with angular momentum $j_i$ 
            represents a phase factor of $(-1)^{j_i-m_i}$.
           }
   \label{t3_ang}
\end{figure}

 To evaluate angular factors it is essential to employ a diagrammatic 
representation which is consistent with the angular momentum coupling 
sequence. Following the conventions of diagrammatic representation of angular 
momentum coupling in ref. \cite{lindgren-86}, the diagram in Fig. 
\ref{t3_ang} is the equivalent angular momentum diagram of the $T_3^{(0)}$ .
This is, however, not the only representation possible, there is another
equivalent and elegant representation described in ref. \cite{porsev-06a}. 
Except for the topology, the two representations require the same number of 
multipoles and should give the same results.

In the representation we have used the diagram is symmetric or invariant under 
the permutation of the vertices. There are four vertices in the angular 
momentum diagram, out of which three involve coupling of angular momentum of 
the spin-orbitals and these are $(j_a, j_p,l_1)$, $(j_b,j_q,l_2)$ and 
$(j_c, j_r, l_3)$. The last one, $(l_1,l_2,l_3)$, involves coupling of the 
multipoles associated with the orbital vertices. Following the angular 
momentum coupling, the vertices must satisfy the triangular conditions
$|j_a-j_p|\leqslant l_1\leqslant (j_a+j_p)$,
$|j_b-j_q|\leqslant l_2\leqslant (j_b+j_q)$,
$|j_c-j_r|\leqslant l_3\leqslant (j_c+j_r)$ and
$|l_1-l_2|\leqslant l_3\leqslant (l_1+l_2)$. Similarly, from the 
parity considerations, the orbital angular momenta must satisfy the
condition that $l_a+l_b+l_c+l_p+l_q+l_r$ is even.


\section{Dipole Polarizability}

 From the second order time-independent perturbation theory, the 
ground state dipole polarizability of a closed-shell atom is 
\begin{equation}
  \alpha = -2  \sum_{I} \frac
  {\langle \Psi_0|\mathbf D|\Psi_I\rangle \langle \Psi_I|\mathbf D|
  \Psi_0\rangle}{E_0 - E_I}, 
\end{equation}
where $|\Psi_I \rangle $ are the intermediate atomic states and $E_I$ is the 
energy of the atomic state. As $\mathbf{D}$ is an odd parity operator, 
$|\Psi_I\rangle$ must be opposite in parity to $|\Psi_0\rangle$. 
For calculations with the RCCSDT wavefunction, there is a subtle issue
depending on how the $T^{(0)}$ equations are implemented and solved. 
By definition, following the linked-cluster theorem, RCCSDT state 
$|\Psi_0\rangle$ consists of only linked and exclusion principle obeying 
(EPO) diagrams. So, in the computational implementations of the RCCSDT 
method, one must ensure that only EPO diagrams are evaluated and included 
in the equations. The electric dipole polarizability 
in terms of the PRCC wavefunction is then
\begin{equation}
  \alpha = -\left ( \langle\tilde{\Psi}_0|\mathbf{D}|\tilde{\Psi}_0\rangle 
            \right ) _{\rm conn},
\end{equation}
where the subscript `conn' represents only connected terms. In practice, 
however, selecting only the EPO diagrams require several conditional 
statements and it is computationally very inefficient. A much faster
implementation is to do an unrestricted evaluation along with
the exclusion principle violating (EPV) diagrams and remove the contributions
at the end of the calculations. Detailed discussions on
different aspects of EPV diagrams in many-body calculations are 
given in ref. \cite{shavitt-09}. So, in general, the terms in the RCCSDT 
equations are computed 
without restrictions on the contracted or internal orbital lines. The computed 
cluster amplitudes and the RCCSDT states, then, have EPV diagrams and for 
future reference define the corresponding RCCSDT and PRCC states as 
$|\Psi'_0\rangle$ and $|\tilde{\Psi}'_0\rangle$, respectively. However, when 
all possible connected diagrams are considered, we can factor the expectation
of $\mathbf{D}$ in terms of the EPV states as product of `connected'
expectation part and normalization factor \cite{Cizek-69}. For the present 
work,  we can write
\begin{equation}
  \langle\tilde{\Psi}'_0|\mathbf{D}|\tilde{\Psi}'_0\rangle
  = \left ( \langle\tilde{\Psi}_0|\mathbf{D}|\tilde{\Psi}_0\rangle 
    \right ) _{\rm conn} \langle\tilde{\Psi}'_0|\tilde{\Psi}'_0\rangle.
\end{equation}
With this definition, the electric dipole polarizability of a closed-shell
atom in terms of the EPV coupled-cluster state is
\begin{equation}
  \alpha = -\frac{\langle\tilde{\Psi}'_0|\mathbf{D}|\tilde{\Psi}'_0\rangle}
           {\langle\tilde{\Psi}'_0|\tilde{\Psi}'_0\rangle}
         = -\left ( \langle\tilde{\Psi}_0|\mathbf{D}|\tilde{\Psi}_0\rangle 
            \right ) _{\rm conn}.
  \label{d_prcc}
\end{equation}
So, it must be mentioned that, in general, the expectation value of an 
operator must be normalized when the condition of EPO is not imposed while 
calculating the cluster amplitudes. The reason is
\begin{equation}
  \left ( \langle\tilde{\Psi}'_0|\mathbf{D}|\tilde{\Psi}'_0\rangle 
            \right ) _{\rm conn} \neq
  \left ( \langle\tilde{\Psi}_0|\mathbf{D}|\tilde{\Psi}_0\rangle 
            \right ) _{\rm conn},
\end{equation}
as $\left ( \langle\tilde{\Psi}'_0|\mathbf{D}|\tilde{\Psi}'_0\rangle 
\right ) _{\rm conn}  $, though connected, retains EPV diagrams subsumed
in the definition of the cluster amplitudes. The definition of $\alpha$ in 
Eq. (\ref{d_prcc}) is what we use in our present and previous works. Since, 
the implementation with EPV diagrams is the usual case, we drop the `prime' 
notation here after and inclusion of EPV diagrams is implied. From the 
definition of $|\tilde{\Psi}_0\rangle$ in Eq. (\ref{psi_tilde}) and based on 
the parity selection rules, only the terms linear in $\lambda$ are nonzero. 
That is,
\begin{equation}
  \alpha = -\frac{\langle \Phi_0|\mathbf{T}^{(1)\dagger}\bar{\mathbf{D}} + 
   \bar{\mathbf{D}}\mathbf{T}^{(1)}|\Phi_0\rangle}{\langle\Psi_0|\Psi_0\rangle},
\end{equation}
where, $\bar{\mathbf{D}} = e^{{T}^{(0)\dagger}}\mathbf{D} e^{T^{(0)}}$, 
represents the unitary transformed electric dipole operator and 
$\langle\Psi_0|\Psi_0\rangle$ is the normalization factor.  From here on, 
it is implicit that expressions with more than one operator involves
contraction and for compact notation, we drop the notation to represent
operator contractions. Retaining the the leading order terms, we obtain
\begin{eqnarray}
 \alpha & \approx & \frac{1}{\cal N}\langle\Phi_0|
     \mathbf{T}_1^{(1)\dagger}\mathbf{D} + \mathbf{D}\mathbf{T}_1^{(1)} 
     + \mathbf{T}_1^{(1)\dagger}\mathbf{D}T_1^{(0)} 
     + T_1^{(0)\dagger}\mathbf{D}\mathbf{T}_1^{(1)}\nonumber \\
    &&  + \mathbf{T}_2^{(1)\dagger}\mathbf{D}T_1^{(0)} 
     + T_1^{(0)\dagger}\mathbf{D}\mathbf{T}_2^{(1)}
     + \mathbf{T}_1^{(1)\dagger}\mathbf{D}T_2^{(0)}\nonumber \\ 
    && + T_2^{(0)\dagger}\mathbf{D}\mathbf{T}_1^{(1)}
     + \mathbf{T}_2^{(1)\dagger}\mathbf{D}T_2^{(0)} 
     + T_2^{(0)\dagger}\mathbf{D}\mathbf{T}_2^{(1)}
     |\Phi_0\rangle, 
 \label{exp_alpha}
\end{eqnarray}
where ${\cal N} = \langle\Phi_0|\exp[T^{(0)\dagger}]\exp[T^{(0)}]
|\Phi_0\rangle$ is the normalization factor, which involves a non-terminating
series of contractions between ${T^{(0)}}^\dagger $ and $T^{(0)} $. However, in 
the present work we use 
${\cal N} \approx \langle\Phi_0|T_1^{(0)\dagger}T_1^{(0)} + 
T_2^{(0)\dagger}T_2^{(0)}|\Phi_0\rangle$. From the above expression of $\alpha$,
an evident advantage of calculation using PRCC theory is the absence of
summation over $|\Psi_I\rangle $. The summation is subsumed in the 
evaluation of the $\mathbf{T}^{(1)}$ in a natural way. This is 
one of the key advantage of using PRCC theory. 

\begin{figure}[h]
  \begin{center}
  \includegraphics[width=6cm]{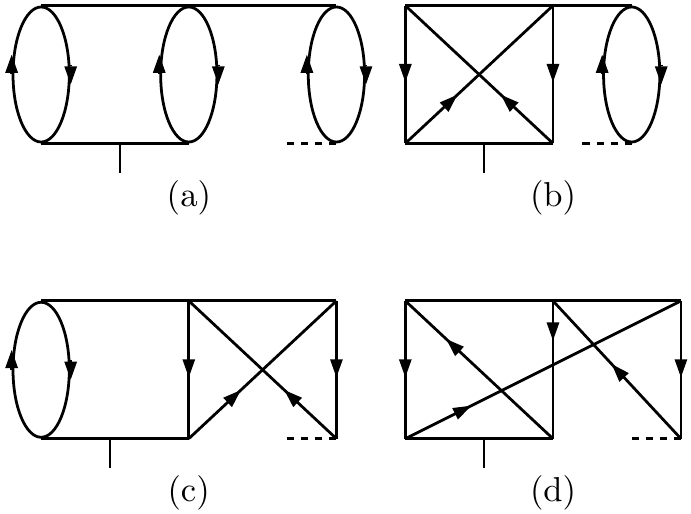}
  \caption{Diagrams of electric dipole polarizability $\alpha$ arising from 
           the term $T_3^{(0)\dagger}\mathbf{D}\mathbf{T}_2^{(1)}$, the
           is the dominant contribution with triple excitation.}
  \label{alpha_triples}
  \end{center}
\end{figure}

 To incorporate the leading order contribution from $T_3^{(0)}$, there are
two possible combinations in the PRCC expression $\alpha$. These are,
${T_3^{(0)}}^{\dagger}\mathbf{D}\mathbf{T}_2^{(1)}$ and hermitian conjugate.
The diagrams which arise from this term, 
$T_3^{(0)\dagger}\mathbf{D}\mathbf{T}_2^{(1)}$, are shown
in Fig. \ref{alpha_triples}.
In the the present work we incorporate these four diagrams to study the 
contribution from $T_3^{(0)}$ in the calculation of $\alpha$. To summarize
this section, we consider all the terms that will arise at the CCSD level and
along with that we consider the diagrams from the contraction of
$T_3^{(0)\dagger}\mathbf{D}\mathbf{T}_2^{(1)}$.


\section{Computational details}

\subsection{Angular integration of diagrams with $T_{3}^{(0)}$}

\begin{figure}[h]
  \begin{center}
  \includegraphics[width=4.5cm]{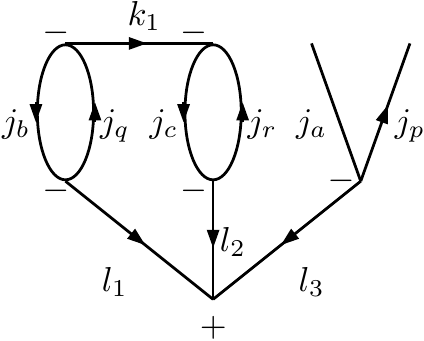}
  \caption{Angular part of the diagram in Fig. \ref{single_triple}(a) 
           represented as angular momentum diagram. The diagram is evaluated
           using standard angular momentum coupling and summation identities
           represented as diagrams. }
  \label{angular_sing}
  \end{center}
\end{figure}

\begin{figure}[h]
  \begin{center}
  \includegraphics[width=8cm]{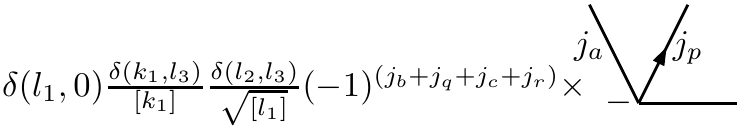}
  \end{center}
  \caption{Expression obtained after angular integration, using diagrammatic 
           techniques, of the diagram in  Fig. \ref{angular_sing}.
          }
  \label{ang_sing_free}
\end{figure}

In this section we discuss the angular integration of the diagrams
arising from $T_{3}^{(0)}$, shown in Fig. \ref{single_triple}, 
\ref{double_triple} and \ref{t3_diag}, in the cluster equations. 
For this we resort to angular momentum diagrams and follow the conventions 
of Lindgren and Morrison \cite{lindgren-86}. We discuss three examples
to describe the notations and conventions we have adopted for 
angular integration. These examples, in particular,  illustrate the
incorporation of the $T_3^{(0)}$ diagrams in the angular momentum diagram 
evaluation. As first example let us consider a relatively simple case, the 
Goldstone diagram in Fig. \ref{single_triple}(a). It is a direct interaction 
diagram, and evaluation does not involve any $3j$- or $6j$- symbols. The 
corresponding angular momentum diagram is shown in Fig. \ref{angular_sing},
where $j_a, j_b, j_c (j_p, j_q, j_r)$ are the total angular momentum of the
core (virtual) orbitals, $k_1$ represents the multipole of the two
electron Coulomb-Breit interaction and $l_i$ are the multipole lines
associated with $T_3^{(0)}$. To simplify and evaluate the diagram we use 
angular momentum diagram identities like bubble removal, and JLV theorems
\cite{lindgren-86}. After simplification, the diagram in 
Fig. \ref{angular_sing} is equivalent to the result shown in 
Fig. \ref{ang_sing_free}.
As can be seen from the figure, the result consists of Kronecker delta 
function, phase factor and angular momentum diagram of the free orbital
lines of the $T_{1}^{(0)}$ operator. Algebraically, the remnant angular 
momentum diagram or the free part in Fig. \ref{ang_sing_free} is equivalent 
to the geometric part in the matrix element defined using the Wigner-Eckart 
theorem, which explicitly depends on the magnetic quantum numbers of the 
initial and final states. The free part, however, is common to all the terms 
and to compute $T^{(0)}$ amplitudes,  we define the RCC equations in 
terms of the reduced matrix elements. 

\begin{figure}[h]
  \begin{center}
  \includegraphics[width=5cm]{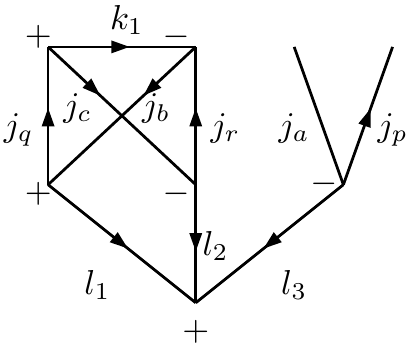}
  \caption{Angular momentum diagram corresponding to the Goldstone diagram
           in Fig. \ref{single_triple}(b). This is a diagram with exchange
           interaction, and in general, the angular momentum diagrams with
           exchange interactions have topologically richer structure.}
  \label{ang_sing_exc}
  \end{center}
\end{figure}

 As the second example, consider the exchange diagram of 
Fig. \ref{single_triple}(a) as shown in Fig. \ref{single_triple}(b). The 
corresponding angular momentum diagram is given in 
Fig. \ref{ang_sing_exc}. Here, we must mention that, in general, the angular 
momentum diagrams of the exchange interaction diagrmas are topologically more 
intricate than the direct interaction diagrams.
For this diagram, after simplification, we obtain the expression shown in 
Fig. \ref{ang_sing_free_exc}. The result, like the previous example,
\begin{figure}[h]
  \begin{center}
  \includegraphics[width=8.5cm]{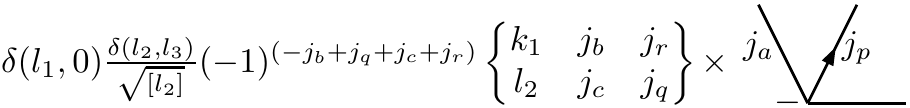}
  \label{ang_sing_free_exc}
  \caption{Expression obtained after angular integration, using diagrammatic
           techniques, of the angular momentum diagram in 
           Fig. \ref{ang_sing_exc}.
  \label{ang_sing_free_exc}
          }
  \end{center}
\end{figure}
consist of Kronecker delta, phase factor and the same free part. However,
unlike in the previous case, there is a $6j$-symbol in the present example.

\begin{figure}[h]
  \begin{center}
  \includegraphics[width=5cm]{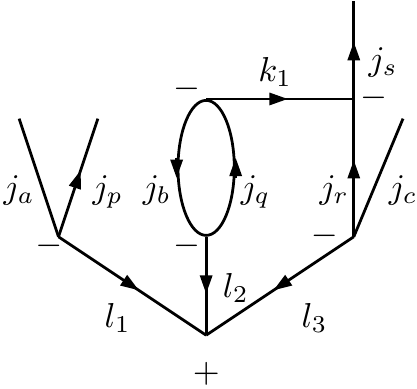}
  \caption{Angular momentum diagram of the Goldstone diagram in 
           Fig. \ref{double_triple}(a) corresponding to the $T_{2}^{(0)}$ 
           cluster amplitude.}
  \label{ang_dbl}
  \end{center}
\end{figure}

\begin{figure}[h]
  \begin{center}
  \includegraphics[width=8cm]{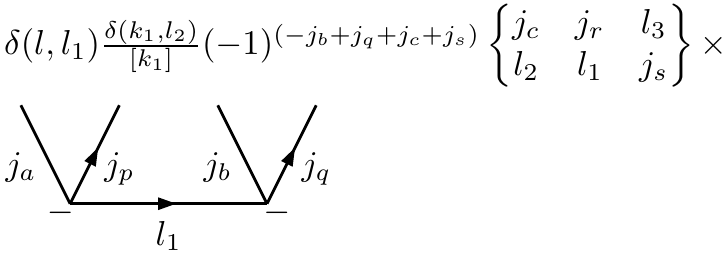}
  \caption{Expression obtained after angular momentum integration, using
           diagrammatic techniques, of the diagram in 
           Fig. \ref{ang_dbl}}.
  \label{ang_dbl_free}
  \end{center}
\end{figure}

 As the last example, consider the diagram in  Fig. \ref{double_triple}(a),
which is a part of the $T_2^{(0)} $ equation. The corresponding angular 
momentum diagram is shown in Fig. \ref{ang_dbl}.
After angular integration, we get the expression shown in 
Fig. \ref{ang_dbl_free}.
The expression is similar in structure to the previous one, however, the 
free part shows the representation adopted for the angular part of $T_2^{(0)}$.
Following the same notations and angular momentum diagram evaluation rules,
the angular integration of the Goldstone diagrams in the $T_3^{(0)}$ cluster
equations are carried out.
\begin{table}[h]
   \caption{The $\alpha_0$ and $\beta$ parameters of the even tempered
            GTO basis used in the present calculations.}
   \label{basis}
   \begin{tabular}{cccccccc}
   \hline
   \hline
     Atom & \multicolumn{2}{c}{$s$} & \multicolumn{2}{c}{$p$} & 
     \multicolumn{2}{c}{$d$}  \\
     & $\alpha_{0}$  & $\beta$ & $\alpha_{0}$ & $\beta$  
     & $\alpha_{0}$  & $\beta$  \\
     \hline
     Mg &\, 0.02950  &\, 1.630 &\, 0.09750 &\, 1.815 &\, 0.00750 &\, 2.710 \\
     Ca &\, 0.02050  &\, 1.970 &\, 0.05250 &\, 1.890 &\, 0.00690 &\, 2.695 \\
     Sr &\, 0.01850  &\, 2.030 &\, 0.04750 &\, 2.070 &\, 0.00910 &\, 2.090 \\
     Ba &\, 0.00925  &\, 2.110 &\, 0.00975 &\, 2.040 &\, 0.00995 &\, 2.010 \\
     Ra &\, 0.00985  &\, 1.990 &\, 0.00925 &\, 1.980 &\, 0.00950 &\, 1.870 \\
     \hline
   \end{tabular}
\end{table}


\subsection{Basis set and nuclear density}
In the present work we use the Dirac-Hartree-Fock Hamiltonian and 
even-tempered Gaussian type orbitals (GTOs) \cite{mohanty-90}. The properties
of the GTO's are described in our previous works
\cite{mani-09, chattopadhyay-12b} and here, we highlight the main 
points. The large component of the Dirac spin-orbitals 
are linear combinations of the Gaussian type functions
\begin{equation}
   g_{\kappa p}^{L}(r) = C^{L}_{\kappa i} r^{n_{\kappa}}e^{-\alpha_{p}r^{2}},
\end{equation}
where $p$ is the index of the Gaussian type function and $C^{L}_{\kappa i}$ 
is the normalization constant. The exponent $\alpha_{p}$ depends on two 
parameters $\alpha_{0}$ and $\beta$, these are related as
$\alpha_{p} = \alpha_{0} \beta^{p-1}$, where $p=0,1\ldots m$ and $m$ is the 
number of the Gaussian type functions. The small components of the 
spin-orbitals are linear combination of $g_{\kappa p}^{S}(r)$, which are 
generated from $g_{\kappa p}^{L}(r)$ through the kinetic balance condition 
\cite{stanton-84}. We compute the GTOs on a grid \cite{chaudhuri-99} with
$V^{\rm N}$ potential and optimize the values of $\alpha_{0}$ and $\beta$ of 
each atom to match the spin-orbital energies and self consistent field (SCF) 
energy obtained from GRASP2K \cite{jonsson-13}. The latter solves the 
Dirac-Hartree-Fock equations numerically and for better convergence we use the 
Hartree-Fock orbitals \cite{fischer-87} as the starting values of GRASP2K. 
The symmetry wise values of the optimized $\alpha_{0}$ and $\beta$ are listed 
in Table. \ref{basis}. Here we must emphasize that the proper optimization
of $\alpha_{0}$ and $\beta$ is important as the quality of
Dirac-Hartree-Fock orbitals depend on these parameters.  
\begin{table}[h]
        \caption{Comparison between GTO and GRASP2K SCF Energies}
        \label{scf}
        \begin{center}
        \begin{tabular}{ldd}
            \hline
            Atom &
              \multicolumn{1}{c}{GTO} &
              \multicolumn{1}{c}{GRASP2K}  \\
            \hline
            Mg & -199.9304   & -199.9351   \\
            Ca & -679.7100   & -679.7102   \\
            Sr & -3178.0789  & -3178.0801  \\
            Ba & -8135.6412  & -8135.6475  \\
            Ra & -25028.0819 & -25028.1072 \\
           \hline
        \end{tabular}
        \end{center}
\end{table}
The SCF energies with the optimized basis set parameters are
given in Table. \ref{scf}. The deviation of the GTO results is largest in the
case of Ra and differs from the GRASP2K result by 0.0253 a.u. The orbital
energies are, however, in excellent agreement. As example, the comparison of 
the orbital energies of Ba and Ra are listed in Table. \ref{orb_en}. 
The largest deviation is observe in the $1s_{1/2}$ orbital energies of Ba 
and Ra. For the outer orbitals the difference in the orbital energies is 
below $10^{-4}$ a.u., which is evident from the table. A similar pattern is 
observed for Mg, Ca and Sr.

 In addition to the basis parameter, the size of the orbital basis set is 
another important factor to obtain accurate results. In the present work, 
we chose the optimal orbital basis size after examining the
convergence of $\alpha$. For example, in the case of Ca we start with a basis 
set size of 95 and increases it up to 165. The value of $\alpha$ with 
Dirac-Coulomb Hamiltonian is listed in Table. \ref{basis_ca}.  
\begin{table}[h]
        \caption{Orbital energies of Ba and Ra obtained from 
                 GRASP2K \cite{jonsson-13} and GTO in atomic units.}
        \label{orb_en}
        \begin{center}
        \begin{tabular}{ldddd}
            \hline
              {Orbital} &
              \multicolumn{2}{c}{Ra} &
              \multicolumn{2}{c}{Ba} \\
              & \multicolumn{1}{c}{\text{GTO}}     & 
                \multicolumn{1}{c}{\text{GRASP2K}} & 
                \multicolumn{1}{c}{\text{GTO}}     & 
                \multicolumn{1}{c}{\text{GRASP2K}}   \\
            \hline
            $1s_{1/2}$ & -3845.8119  & -3845.8206  & -1383.8341 & -1383.8358 \\
            $2s_{1/2}$ & -712.6594   & -712.6607   & -222.5774  & -222.5777  \\
            $2p_{1/2}$ & -685.0966   & -685.0966   & -209.0878  & -209.0881  \\
            $2p_{3/2}$ & -571.9348   & -571.9345   & -195.0103  & -195.0103  \\
            $3s_{1/2}$ & -179.6860   & -179.6863   & -48.6517   & -48.6517   \\
            $3p_{1/2}$ & -167.2343   & -167.2343   & -42.9566   & -42.9566   \\
            $3p_{3/2}$ & -141.1856   & -141.1855   & -40.1674   & -40.1673   \\
            $3d_{3/2}$ & -121.3160   & -121.3158   & -30.2980   & -30.2979   \\
            $3d_{5/2}$ & -115.9592   & -115.9590   & -29.7120   & -29.7119   \\
            $4s_{1/2}$ & -45.6825    & -45.6827    & -10.2572   & -10.2572   \\
            $4p_{1/2}$ & -40.1160    & -40.1161    & -8.0991    & -8.0991    \\
            $4p_{3/2}$ & -33.3984    & -33.3984    & -7.5132    & -7.5132    \\
            $4d_{3/2}$ & -24.4189    & -24.4188    & -3.9137    & -3.9135    \\
            $4d_{5/2}$ & -23.1605    & -23.1604    & -3.8126    & -3.8126    \\
            $4f_{5/2}$ & -11.3559    & -11.3560    &            &            \\
            $4f_{7/2}$ & -11.0465    & -11.0465    &            &            \\
            $5s_{1/2}$ & -10.0018    & -10.0020    & -1.6035    & -1.6035    \\
            $5p_{1/2}$ & -7.8439     & -7.8440     & -0.9564    & -0.9564    \\
            $5p_{3/2}$ & -6.3733     & -6.3734     & -0.8727    & -0.8727    \\
            $5d_{3/2}$ & -3.1177     & -3.1177     &            &            \\
            $5d_{5/2}$ & -2.9028     & -2.9028     &            &            \\
            $6s_{1/2}$ & -1.6247     & -1.6247     & -0.1632    & -0.1632    \\
            $6p_{1/2}$ & -0.9740     & -0.9740     &            &            \\
            $6p_{3/2}$ & -0.7406     & -0.7406     &            &            \\
            $7s_{1/2}$ & -0.1662     & -0.1662     &            &            \\
           \hline
        \end{tabular}
        \end{center}
\end{table}
It is evident from the table that, the optimal orbital basis size at which
$\alpha$ saturates to the level of $10^{-2}$ a.u. is 137. As we increase 
beyond this, the change in $\alpha$ is $\leqslant 10^{-3}$ a.u. So, we use the 
orbital basis size 137 for further computations with the DCB Hamiltonian and
to study the contributions from $T_3^{(0)}$.
\begin{table}[h]
  \caption{Convergence pattern of $\alpha$ (Ca) as a function of
           the basis set size.}
  \label{basis_ca}
  \begin{tabular}{lcc}
      \hline
      No. of orbitals & Basis size & $\alpha $   \\
      \hline
      95  & $(13s, 11p, 9d, 7f, 7g, 7h)     $ & 163.74  \\
      113 & $(15s, 13p, 11d, 9f, 9g, 7h)    $ & 163.55  \\
      127 & $(17s, 15p, 11d, 11f, 9g, 9h)   $ & 163.53  \\
      137 & $(19s, 15p, 13d, 11f, 11g, 9h)  $ & 163.52  \\
      147 & $(21s, 17p, 17d, 13f, 11g, 11h) $ & 163.52  \\
      165 & $(23s, 19p, 15d, 13f, 13g, 11h) $ & 163.52  \\
      \hline
  \end{tabular}
\end{table}

  In the present work we use the finite size Fermi density distribution of 
the nucleus
\begin{equation}
   \rho_{\rm nuc}(r) = \frac{\rho_0}{1 + e^{(r-c)/a}},
\end{equation}
where, $a=t 4\ln(3)$. The parameter $c$ is the half charge radius so that 
$\rho_{\rm nuc}(c) = {\rho_0}/{2}$ and $t$ is the skin thickness. In one
of our earlier works \cite{chattopadhyay-13b} we had examined the vacuum 
polarization corrections to the orbital energies and atomic properties. 
In the present work, we neglect the vacuum polarization effects as
we find the corrections in the properties of neutral alkaline atoms are
very small. The PRCC equations are solved iteratively using Jacobi method, 
we have chosen this method as it is easily parallelizable. 
The method, however, is slow to converge. So, we use direct inversion in the 
iterated subspace (DIIS) \cite{pulay-80} to accelerate the convergence.


\subsection{Convergence criteria}

The computation of $\alpha$, as described earlier, involves several steps: 
the generation of orbital basis set; computation of the RCC cluster
amplitudes ($T^{(0)}$ ); and computation of PRCC cluster amplitudes 
($\mathbf{T}^{(1)}$). Each of these are iterative in nature and involves the
choice of a convergence parameter $\epsilon$. In the SCF computations to 
generate the orbital basis set, $\epsilon$ is the maximum change in the 
orbital between two consecutive iterations, and in the computations of 
$T^{(0)}$ and $\mathbf{T}^{(1)}$, it is the average 
change in the cluster amplitudes between consecutive iterations. 
In all of these we set choose $\epsilon \leqslant10^{-6}$ and it is 
computationally manageable with single and double excitation approximation
in the coupled-cluster computation. However, with the the inclusion of the 
triple excitations, we choose $\epsilon\leqslant 10^{-5}$. This choice is
compelled by the computational requirements and the effect on the value
of $\alpha$ is  below $10^{-4}$ a.u. To elaborate further,
when $T_3^{(0)}$ is included, the number of cluster amplitudes increases by 
several order of magnitudes and each iteration requires thousands of compute 
hours.  For example, in Ca, consider the orbital basis with 137 orbitals, the 
optimal size to obtain converged value of $\alpha$. With this basis set,
number of the $T_1^{(0)}$ and $T_2^{(0)}$ cluster amplitudes are
$\approx 10^2$ and $\approx 10^5$, respectively. On the other hand, the
number of $T_3^{(0)}$ cluster amplitudes is $\approx 10^{10}$. So, it 
involves massive computational operations and in addition,
due to the large number we observe slower convergence. We find that even
with the reduction in the number of  $T_3^{(0)}$ to $\approx 10^{8}$, 
based on energy considerations, solving the cluster equations takes weeks
on multi-node cluster computers. So, decreasing $\epsilon$ by an order
magnitude to $10^{-5}$ reduces a few iterations, and we can solve the 
cluster equations in about {\em three weeks} using 64 processors on a 
cluster computer.


\subsection{Computational issues related to $T_3^{(0)}$}

As discussed in the previous section, to compute the $T_3^{(0)}$ cluster
amplitudes within reasonable time, we chose dominant $T_3^{(0)}$ amplitudes
based on the orbital energies. For this consider the perturbative 
approximation of the $T_3^{(0)}$ cluster amplitude
\begin{equation}
  t_{abc}^{pqr} \approx \frac{1}{\Delta E} \left ( 
       \sum_s t_{ab}^{ps}g_{sc}^{qr} + \sum_d t_{ad}^{pq}g_{bc}^{dr} \right ),
\end{equation}
where, as defined earlier $t_{ab}^{pq}$ is the $T_2^{(0)}$ cluster amplitude,
$g_{ij}^{kl} = \langle{kl}|1/r_{12} + g^{B}(r_{12})|{ij} \rangle$, and the
energy denominator is
\begin{equation}
   \Delta{E} = {(\varepsilon_{p} + \varepsilon_{q} + \varepsilon_{r} - 
   \varepsilon_{a} - \varepsilon_{b} -\varepsilon_{c})}.
\end{equation}
In the above expression, $\varepsilon_i$ is the single particle energy of the 
$i$th orbital. Since, $ t_{abc}^{pqr} \propto 1/\Delta{E}$, the dominant
contributions arise from small $\Delta{E}$ and for the present work we select
the orbitals such that $\Delta{E}\leqslant 100$ a.u. Here, we have set the 
selection criterion based on $\Delta{E}$ as the numerators
$t_{ab}^{ps}g_{sc}^{qr}$ and $t_{ad}^{pq}g_{bc}^{dr}$ are, in general, smaller
than one when $\Delta{E}>10$. 

  There is an important consideration associated with the $T_3^{(0)}$
cluster amplitudes during the computation. This is related to the maximum
occupancy of the $s_{1/2}$ and $p_{1/2}$ orbitals, which is two, to avoid
violation of Pauli's exclusion principle. During 
computations we ensure that not all the $a$, $b$ and $c$ in $t_{abc}^{pqr}$ 
are associated with the same $s_{1/2}$ or $p_{1/2}$ sub-shells. Similar 
consideration must be made for the $p$, $q$ and $r$. Implementing this 
explicitly as a selection rule, however, leads to larger computational
operations as $t_{abc}^{pqr}$ contributes to single, double and triple
excitation cluster amplitudes. A more efficient and faster scheme is to
allow an un-constrained computation of the terms in the cluster equations,
including the terms with triple occupancy of $s_{1/2}$ and $p_{1/2}$
sub-shells. We then, subtract these terms just before computing the
new values of the cluster amplitudes. This method of selected subtraction
speeds up computation as it avoids global implementation of a selection rule
requiring multiple conditional statements.
\begin{table}[h]
  \caption{Static dipole polarizability $\alpha$ of alkaline-Earth-metal  
           atoms and the values are in atomic units. The results from this
           work identified with C and B within parenthesis indicate the 
           use of Dirac-Coulomb and Dirac-Coulomb-Breit Hamiltonian, 
           respectively.}
  \label{pol_alkaline}
  \begin{center}
  \begin{tabular}{ldccc}
    \hline
    \multicolumn{1}{c}{Atom}      & \multicolumn{1}{c}{This work}      &
    \multicolumn{1}{c}{Method}    & \multicolumn{1}{c}{Previous Works} &
    \multicolumn{1}{c}{Method}                              \\
    \hline
    \hline
    $\rm{Mg}$ & 71.64  & PRCC(C)  & 71.35     \footnotemark[1] & CICP       \\
              & 70.76  & PRCC(B)  & 70.90     \footnotemark[2] &            \\
              &        &          & 72.54     \footnotemark[3] & RCCSD      \\
              &        &          & 71.33     \footnotemark[4] & RCI + MBPT \\
              &        &          & 71.5(3.1) \footnotemark[5] & Expt.      \\
    $\rm{Ca}$ & 163.52 & PRCC(C)  & 158.00    \footnotemark[6] & RCCSDT     \\
              & 160.77 & PRCC(B)  & 152.0     \footnotemark[7] & RCCSDT     \\
              &        &          & 157.03    \footnotemark[3] & RCCSD      \\
              &        &          & 159.0     \footnotemark[4] & RCI + MBPT \\
              &        &          & 169(17)   \footnotemark[8] & Expt.      \\
    $\rm{Sr}$ & 192.40 & PRCC(C)  & 198.85    \footnotemark[6] & RCCSDT     \\
              & 190.82 & PRCC(B)  & 190       \footnotemark[7] & RCCSDT     \\
              &        &          & 186.98    \footnotemark[3] & RCCSD      \\
              &        &          & 202.0     \footnotemark[4] & RCI + MBPT \\
              &        &          & 186(15)   \footnotemark[9] & Expt.      \\
    $\rm{Ba}$ & 278.24 & PRCC(C)  & 273.9     \footnotemark[6] & RCCSDT     \\
              & 274.68 & PRCC(B)  & 275.5     \footnotemark[10] & RCCSDT    \\
              &        &          & 268.19(7.28)\footnotemark[11]& RCCSD    \\
              &        &          & 272.1     \footnotemark[4] & RCI + MBPT \\
              &        &          & 268(22)   \footnotemark[9] & Expt.      \\
    $\rm{Ra}$ & 243.52 & PRCC(C)  & 248.56    \footnotemark[6] & RCCSDT     \\
              & 242.42 & PRCC(B)  &                            &            \\
    \hline
    \hline
  \end{tabular}
  \end{center}
\footnotetext[1]{Reference\cite{mitroy-03}.}
\footnotetext[2]{Reference\cite{hamonou-08}.}
\footnotetext[3]{Reference\cite{singh-13}.}
\footnotetext[4]{Reference\cite{porsev-06c}.}
\footnotetext[5]{Reference\cite{lundin-73}.}
\footnotetext[6]{Reference\cite{lim-04}.}
\footnotetext[7]{Reference\cite{sadlej-91}.}
\footnotetext[8]{Reference\cite{miller-76}.}
\footnotetext[9]{Reference\cite{schwartz-74}.}
\footnotetext[10]{Reference\cite{sch-07}.}
\footnotetext[11]{Reference\cite{sahoo-08}.}
\end{table}


\section{Results and Discussions}

In the present work, we use the expression of $\alpha$ in 
Eq. (\ref{exp_alpha}), which consider terms up to second order in
cluster operators. This is a natural choice as the contributions from the 
higher order terms are very small $\approx 10^{-4}$ a.u. It is important 
to mention that unlike the closed-shell noble gas atoms and one-valence 
alkali atoms, the alkaline earth metal atoms are two-valence systems and 
have more complex structure. Further more, the uncertainties in the 
experimental results are also large. So, it is important to identify and 
quantify the theoretical uncertainties in the computations. 
In table \ref{pol_alkaline} we list the $\alpha$ of alkaline-Earth metal
atoms Mg, Ca, Sr, Ba and Ra computed using the PRCC expression of $\alpha$. 
In the second column of the table we give our results with 
two different sets of calculations : one is with the Dirac-Coulomb (DC) 
Hamiltonian, and the other is with the Dirac-Coulomb-Breit (DCB) Hamiltonian. 
For a systematic comparison we also list the previous theoretical and 
experimental results. 
\begin{table}[h]
    \caption{Contribution to $\alpha $ from different terms and their
             hermitian conjugates in the PRCC theory.}
    \label{result_prcc}
    \begin{center}
    \begin{tabular}{lddddd}
        \hline
        Terms + h.c. & \multicolumn{1}{r}{$\rm{Mg}$}
        & \multicolumn{1}{r}{$\rm{Ca}$}
        & \multicolumn{1}{r}{$\rm{Sr}$}
        & \multicolumn{1}{r}{$\rm{Ba}$}
        & \multicolumn{1}{r}{$\rm{Ra}$}  \\
        \hline
        $\mathbf{T}_1^{(1)\dagger}\mathbf{D} $
        & 75.790   &  185.852  &  224.452  &  342.336   &  311.590     \\
        $\mathbf{T}_1{^{(1)\dagger}}\mathbf{D}T_2^{(0)} $
        & -2.400   & -7.254    & -9.130    & -18.674    & -16.462      \\
        $\mathbf{T}_2{^{(1)\dagger}}\mathbf{D}T_2^{(0)} $
        &  2.706   &  10.004   &  12.926   &  30.348    &  23.506      \\
        $\mathbf{T}_1{^{(1)\dagger}}\mathbf{D}T_1^{(0)} $
        & -2.056   & -9.926    & -12.522   & -29.008    & -27.200      \\
        $\mathbf{T}_2{^{(1)\dagger}}\mathbf{D}T_1^{(0)} $
        &  0.184   &  0.906    &  1.236    &  3.238     &   3.104      \\
        Normalization & 1.049  & 1.117   &  1.137  & 1.195   & 1.215   \\
        Total         & 70.757 & 160.771 & 190.820 & 274.678 & 242.418 \\
        \hline
    \end{tabular}
    \end{center}
\end{table}

  For a more detailed study, we examine the contributions from  each of
the terms in the Eq. (\ref{exp_alpha}) and these are listed in
Table.  \ref{result_prcc}. In all the cases, the leading order (LO)
term is $\mathbf{T}_1^{(1)\dagger}\mathbf{D} $ + 
h.c. This is natural as these terms subsume the contributions from the
Dirac-Fock and RPA effects. Further more, the contribution from the LO term
exceeds the final value of $\alpha$ and a similar trend was observed in our 
earlier work on noble gas atoms \cite{chattopadhyay-12a, chattopadhyay-12b}
as well as the alkali-metal \cite{chattopadhyay-13a} and 
alkaline-earth-metal \cite{chattopadhyay-13b} ions. To examine the contribution
from $\mathbf{T}_1^{(1)\dagger}\mathbf{D} $ in finer detail, we separate out
and list the contribution from each of the occupied orbitals to this term in 
the Table. \ref{result_t1d}. In all the cases, the valence orbital  
surpasses contributions from the other occupied orbitals by orders magnitude.
More importantly, in all the cases, the contributions from the inner occupied 
orbitals are opposite to that of the valence orbitals.

On examining the trend in the next to leading order (NLO) term, there is a
significant departure from the general trend observed in  our previous works on
noble gas atoms \cite{chattopadhyay-12a, chattopadhyay-12b},
alkali-metal ions \cite{chattopadhyay-13a} and 
alkaline-earth-metal ions \cite{chattopadhyay-13b}. In these systems the 
NLO term is $\mathbf{T}_1{^{(1)\dagger}}\mathbf{D}T_2^{(0)} $ 
and contribution is opposite in phase to the LO term. However, in neutral 
alkaline atoms, as evident from Table. \ref{pol_alkaline}, the NLO term is 
$\mathbf{T}_2{^{(1)\dagger}}\mathbf{D}T_2^{(0)}$ and the contribution has the
same phase as the LO term. Another important point is, the NLO and the other
important terms,
$\mathbf{T}_1{^{(1)\dagger}}\mathbf{D}T_1^{(0)}$ and 
$\mathbf{T}_1{^{(1)\dagger}}\mathbf{D}T_2^{(0)}$ have much larger contributions
compared to the systems studied earlier. For better presentation of the 
detailed analysis of the results, we separate the discussion
into three groups: Mg and Ca, Sr and Ba, and Ra.
\begin{table}[h]
    \caption{Four leading contributions to
        $\{ \mathbf{T}_1^{(1)\dagger}\mathbf{D}  \}$ to $\alpha $
        in terms of the core spin-orbitals. }
    \label{result_t1d}
    \begin{center}
    \begin{tabular}{rrr}
        \hline
          \multicolumn{1}{c}{Mg} & \multicolumn{1}{c}{Ca}
        & \multicolumn{1}{c}{Sr}  \\ \hline
        38.308 (3$s_{1/2}$) & 95.257 (4$s_{1/2}$)  & 114.342 (5$s_{1/2}$) \\
       -0.032  (2$p_{3/2}$) & -0.529 (3$p_{3/2}$)  & -0.715  (4$p_{3/2}$) \\
       -0.016  (2$p_{1/2}$) & -0.266 (3$p_{1/2}$)  & -0.336  (4$p_{1/2}$) \\
       -0.007  (2$s_{1/2}$) & -0.012 (3$s_{1/2}$)  & -0.012  (3$d_{5/2}$) \\
        \hline
       \multicolumn{1}{c}{Ba}   & \multicolumn{1}{c}{Ra}
      &  \\  \hline
          175.500 (6$s_{1/2}$) & 157.974 (7$s_{1/2}$) & \\
          -1.400  (5$p_{3/2}$) & -0.760  (6$p_{3/2}$) & \\
          -0.567  (5$p_{1/2}$) & -0.202  (6$p_{1/2}$) & \\
          -0.072  (4$d_{5/2}$) & -0.101  (5$d_{5/2}$) & \\ \hline
    \end{tabular}
    \end{center}
\end{table}


\subsection{Mg and Ca}

The results for $\alpha$ of Mg computed with the DC Hamiltonian is in 
excellent agreement with the experimental value. With $ H^{\rm DCB}$,
the value of $\alpha$ is lowered by $-0.88$ a.u., which is not negligible.  
This is in contrast to the case of noble gas atoms, for which we observed
an increase in $\alpha$ when Breit interaction is  included
\cite{chattopadhyay-12b}. This is due to the different orbital angular
momentum of the valence shell, which has the leading order contribution
to $\alpha$. More precisely, the $3s$ orbital in Mg is radially contracted 
when Breit interaction is included, where as in the noble gas atoms the
valence orbital ( $np_{3/2}$ ) is radially dilated with Breit interaction.
Our results for Mg is in good agreement with the previous theoretical 
values, including the results from a relativistic coupled-cluster computations
\cite{singh-13}. In particular, the DC result of Mg is in very good 
agreement with the recommended value of a previous theoretical work 
\cite{porsev-06c} as well as the results from configuration interaction with 
semi empirical core potential model \cite{mitroy-03}, however, the DCB result 
is 0.8\% lower than the recommended value. Among the various terms in the 
PRCC expression of $\alpha$, as described earlier, the LO term is
$\mathbf{T}_1{^{(1)\dagger}}\mathbf{D}$. On examining further, based on the 
values listed in Table. \ref{result_t1d}, almost the 
total value of $\alpha$ arises from the core orbital $3s$. The contribution
from the next core shell $2p_{3/2}$, in terms of energy, is $\approx 0.08$\%,
which is negligible. At a finer level, the five dominant ${\tau}_a^{p}$ 
in $\mathbf{T}_1{^{(1)\dagger}}\mathbf{D}$ are listed in Table. \ref{t1d_contr}.
According to the table, the dominant contributions arise from the 
$\tau_{3s_{1/2}}^{3p_{3/2}} $ and $\tau_{3s_{1/2}}^{3p_{1/2}}$ cluster 
amplitudes.
\begin{table}[h]
  \caption{Orbitals contribution from
           $\mathbf{T}_1{^{(1)\dagger}} \mathbf{D}$ to $\alpha$
           of Mg and Ca}
    \label{t1d_contr}
    \begin{center}
    \begin{tabular}{dcdc}
       \hline
       \multicolumn{2}{c}{Mg}  & \multicolumn{2}{c}{Ca} \\
       \hline
        25.475  & $(3s_{1/2}, 3p_{3/2})$  &   62.072 & $(4s_{1/2}, 4p_{3/2})$ \\
        12.780  & $(3s_{1/2}, 3p_{1/2})$  &   33.159 & $(4s_{1/2}, 4p_{1/2})$ \\
         0.032  & $(3s_{1/2}, 4p_{3/2})$  &  -0.229  & $(3p_{3/2}, 6d_{3/2})$ \\
         0.016  & $(3s_{1/2}, 4p_{1/2})$  &  -0.138  & $(3p_{3/2}, 5d_{3/2})$ \\
     \hline
     \multicolumn{2}{c}{Sr}  & \multicolumn{2}{c}{Ba}
                    \\ \hline
        73.370  & $(5s_{1/2}, 5p_{3/2})$  &   76.628 & $(6s_{1/2}, 6p_{3/2})$\\
        40.975  & $(5s_{1/2}, 5p_{1/2})$  &   45.697 & $(6s_{1/2}, 6p_{1/2})$\\
       -0.236   & $(4p_{3/2}, 6d_{5/2})$  &   32.977 & $(6s_{1/2}, 7p_{3/2})$\\
       -0.176   & $(4p_{3/2}, 7d_{5/2})$  &   16.538 & $(6s_{1/2}, 7p_{1/2})$\\
     \hline
       \multicolumn{2}{c}{Ra} && \\ \hline
        48.128 & $(7s_{1/2}, 7p_{3/2})$  && \\
        45.844 & $(7s_{1/2}, 7p_{1/2})$  && \\
        34.735 & $(7s_{1/2}, 8p_{3/2})$  && \\
        22.021 & $(7s_{1/2}, 8p_{1/2})$  && \\ \hline
    \end{tabular}
    \end{center}
\end{table}

  For Ca, there are three previous results based on RCC theory 
\cite{sadlej-91,lim-04,singh-13}, however, there is a variation among
the results. Compared to these previous results, our result of $\alpha$ with 
the DC Hamiltonian is on the higher side, but with the inclusion of Breit 
interaction our result is in good agreement with the values given in 
ref. \cite{lim-04} and \cite{singh-13}. In particular, our result is 
1.7\% higher than the RCCSDT result  obtained with finite 
field method \cite{lim-04} with scalar relativistic Douglas-Kroll Hamiltonian,
which is a different approach compared to our method. Further more, our 
result is 1.1\% higher that result from calculations with hybrid RCI+MBPT 
method \cite{porsev-06c}. 

 When compared to Mg, in Ca we notice a large increase in 
the value of $\alpha$. This may be understood in terms of the difference in 
the properties of the valence orbitals generated with the GTO or numerically
using GRASP2K. In Mg, the valence orbital $3s$ has an energy of $-0.255$ 
hartree and average radial extent $\langle r\rangle=3.252 \text{a}_0$. For Ca, 
the valence orbital is $4s$, and the energy and $\langle r\rangle$ are $-0.198$ 
hartree and $4.191 \text{a}_0$, respectively. Thus, the valence 
orbital of Ca is $\approx$ 30\% larger in size, and as $\alpha \propto r^2$, 
we can expect larger $\alpha$. This is reflected in the dominant 
$\mathbf{T}_1^{(1)}$ cluster amplitude which contributes to the leading 
order term in $\alpha$, namely $\mathbf{T}_1{^{(1)\dagger}}\mathbf{D}$ and it's
complex conjugate. In Ca, the contribution from $\tau_{4s_{1/2}}^{4p_{3/2}}$ 
to $\alpha$, $\mathbf{T}_1{^{(1)\dagger}}\mathbf{D}$ is $62.072$. This is 
$\approx 2.5$ times larger than the dominant contribution in Mg arising from 
the $\tau_{3s_{1/2}}^{3p_{3/2}}$ cluster amplitude. A similar trend is observe 
in the second most dominant cluster amplitude.


\subsection{Sr and Ba}

  In the case of Sr too, like in Ca, there is a variation in the previous 
theoretical results obtained from RCC theory 
\cite{sadlej-91,lim-04,singh-13}. However, unlike in Ca, our result 
with DC Hamiltonian lies between the previous RCC results, and with Breit 
interaction our result is almost an exact match with the RCC result
\cite{sadlej-91} using Cowan-Griffin approximation \cite{cowan-76}. This is 
as expected since the relativistic corrections, though important, are not
very large for neutral atoms like Sr with nuclear charge $Z=38$ as $Z\alpha<1$
(here, $\alpha$ is the fine structure constant). On a closer inspection, the 
Breit interaction contribution to $\alpha$ is $-1.58$ a.u., which is 
$\approx 0.8$\% of the total value and this is marginally lower than in Ca. 
This could be due to the screening effect from the electrons in the 
diffused $3d$ shell. 

  The remaining two atoms in the group, Ba and Ra with nuclear charges $56$ 
and $88$, respectively, are good candidate atoms to examine the relativistic 
implementation of coupled-cluster theories in detail. In particular, the 
PRCC theory we have developed for properties calculations with an additional
perturbation. For Ba, three of the previous results
\cite{lim-04,sch-07,sahoo-08} are based on RCC theory, and our results are 
consistent the previous results in ref. \cite{lim-04,sch-07}. Referring to the
third RCC work \cite{sahoo-08}, accounting for the theoretical uncertainty 
reported, we may consider our results consistent with the values reported in 
that work. However, a more detailed comparison is non-trivial as the 
normalization factor reported in ref. \cite{sahoo-08}  is less than unity.
Perhaps, this is on account of the scheme adopted to implement the relativistic 
coupled-cluster theory in their work. 
On inclusion of Breit interaction, using DCB Hamiltonian, the value of 
$\alpha$ is  reduced by $-3.56$ a.u., which is $\approx 1.3$\% of the value
obtained with DC Hamiltonian. Our result with DCB Hamiltonian is in very
good agreement with the previous theoretical result obtained from RCCSDT
\cite{lim-04}, more precisely, our result is 0.3\% lower than the value 
reported in ref. \cite{lim-04}. On the other hand, our result is  0.9\%
higher than the result from the hybrid RCI+MBPT \cite{porsev-06c}.

 The relative change in $\alpha$ as we compare the value of Ba to Sr is not 
very remarkable. However, on closer examination, there is a pronounced change
in the identity of the dominant $\tau_a^p$ contributing to the leading PRCC
term $\mathbf{T}_1{^{(1)\dagger}}\mathbf{D}$. Like in the other atoms 
discussed so far, as evident from Table. \ref{t1d_contr}, the two leading 
contributions arise from the cluster amplitudes with the outermost $s$ and 
$p$ orbitals. More precisely, these are the cluster amplitudes 
$\tau_{6s_{1/2}}^{6p_{3/2}} $ and $\tau_{6s_{1/2}}^{6p_{1/2}}$. The 
contributions from the next two dominant cluster amplitudes 
$\tau_{6s_{1/2}}^{7p_{3/2}}$  and $\tau_{6s_{1/2}}^{7p_{1/2}}$ are 
$\approx$43\% and $\approx$22\% of the most dominant 
($\tau_{6s_{1/2}}^{6p_{3/2}}$) contribution, respectively. This is very 
different from the trend observed in Mg, Ca and Sr, where the third dominant 
contribution is below 1\% of the most dominant contribution. Further more, in 
the lighter atoms Ca and Sr, the third dominant contribution arises from the 
cluster amplitude of the form $\tau_{3p_{3/2}}^{5d_{3/2}}$ and 
$\tau_{4p_{3/2}}^{6d_{5/2}}$, respectively.  This could be on account
of the relativistic contraction of the $s$ and $p$ orbitals. These results
indicate that the choice of the basis set is very important to obtain 
reliable results in Ba.


\subsection{Ra}

We consider the case of Ra as the most important among the alkaline-earth
metal atoms for the present study, and it is also the most
challenging. The primary reasons are: importance of the relativistic effects; 
and absence of experimental data. Considering that $Z$ is 88 and 
$Z\alpha\approx 0.64$ (where $\alpha$ is the fine structure constant), it is 
absolutely essential to employ a relativistic description. For Ra, there is
one previous theoretical result based on RCCSDT using 
scalar relativistic Douglas-Kroll operator $\alpha$ \cite{lim-04}. As evident
from Table. \ref{pol_alkaline}, our result of $\alpha$ with DC Hamiltonian is 
$\approx$2\% lower than the result in ref. \cite{lim-04}. On including
the Breit interaction the difference increases to 2.5\%. An important
observation is, compared to the case of Ba the value of $\alpha$ for Ra is 
$\approx$ 11.7\% lower. A similar trend is observed in the results
reported in ref. \cite{lim-04}, in their case the $\alpha$ of Ra is 
$\approx$9.3\% lower than Ba. This reduction is on account of the radial 
contraction of the $7s$ orbital, the valence orbital of Ra, due to the 
relativistic effects. This is evident from the $\langle r\rangle$ of the 
orbitals generated either with GTO or numerically using GRASP2K. In Ra, the 
$\langle r\rangle$ of the valence orbital $7s$ is 5.04$a_0$, this 
is lower than the value of 5.06$a_0$ for the valence orbital $6s$ in Ba. 

 To identify and as well as quantify the relative change from the $\alpha$
of Ba, let us examine the dominant $\tau_a^p$ which contribute to the 
leading order term in PRCC $\mathbf{T}_1^{(1)\dagger}\mathbf{D}$. From the 
values listed in Table \ref{t1d_contr}, in Ra the most dominant contribution
of 48.128 a.u. arises from the $\tau_{7s_{1/2}}^{7p_{3/2}}$ cluster amplitude. 
However, compared to the most dominant contribution of 76.628 a.u. in Ba, 
arising from the $\tau_{6s_{1/2}}^{6p_{3/2}}$ cluster amplitude, it is 
$\approx$37.2\% smaller. For the next three dominant $\tau_a^p$ cluster 
amplitudes, the contributions in Ra are on par with those of Ba or slightly.
So, most of the changes or reduction in $\alpha$ can be attributed to the 
lower contribution from the dominant cluster amplitude in the LO term 
$\mathbf{T}_1^{(1)\dagger}\mathbf{D}$. Coming to the contribution
from the Breit interaction, after it's inclusion the value of $\alpha$ is 
1.10 a.u. less than the result with DC Hamiltonian.


\subsection{$T_3^{(0)}$ Contribution to $\alpha$}

Here we examine the contribution from $T_3^{(0)} $  to $\alpha$. For this
we compute the most dominant contribution from  $T_3^{(0)} $. In general,
as we have shown earlier, the leading order contribution arises from the terms
involving  $T_1^{1}$ cluster operators. However, with $T_3^{(0)} $ only the
terms with structural radiation diagrams can have non-zero contribution 
involving $T_1^{(1)}$. In general, the contributions from the structural 
radiation diagrams are negligible and not included in the present work.
So, the dominant contribution from $T_3^{(0)} $ arises from the term
${T}_3^{(0)\dagger}\mathbf{D}\mathbf{T}_2^{(1)}$. The diagrams 
of $\alpha$ arising from this term are shown in Fig. \ref{alpha_triples}. 
The computation with $T_3^{(0)}$ is restricted to the cluster amplitudes
involving the outer core orbitals and low-lying virtual orbitals. This,
as mentioned earlier, is due to the limitations of the computational 
resources. Even then, it takes close to a month to solve the $T_3^{(0)}$ 
on cluster computers for lighter atoms.  So, in this work we report the 
results of $\alpha$ for only Mg and Ca with $T_3^{(0)}$. The contribution is 
small and it is $0.0016$ a.u. and $0.0317$ a.u for Mg and Ca, respectively. 
This may be on account of considering only the triple excitations of RCC 
amplitudes.  On the other hand the contribution from $\mathbf{T}_3^{(1)}$ may 
not be small.  We are in the process of developing a theory to incorporate 
$\mathbf{T}_3^{(1)}$, but computational implementation is non-trivial as the 
representation now involves four multipole operators.


\subsection{Uncertainty estimates}

To estimate the uncertainties associated with the present results, we have 
identified some important sources. These are associated with the various 
approximations at different levels of the computations. The first source of 
uncertainty in our computations is associated the truncation of  the orbital 
basis set. To reduce this uncertainty we do a series of computations and
as we discussed earlier, identify the optimal basis set size. So it is safe 
to neglect this uncertainty. The second source of uncertainty is, as we
consider upto $T_3^{(0)}$ in the RCC theory, the contribution from
the $T_4^{(0)}$ and other higher order excitations is a source of
uncertainty. We have shown the contribution from the triple excitations
is $\approx 10^{-2}$ for Ca and we expect a similar pattern for other 
heavy atoms, Sr, Ba and Ra. So the uncertainty associated with the
$T_4^{(0)}$ and other higher order excitations can be ignored. The third
source of uncertainty arises from the contribution of $\mathbf{T}_3^{(1)}$ 
and higher order excitation cluster operators in PRCC theory. Considering 
the results from the finite field calculations in ref. \cite{lim-04}, the 
average contribution from $T_3^{(0)}$ is $\approx 2.8$\%. So, in the 
present work we may take this as the upper bound on the uncertainty 
associated with the truncation of cluster operators in PRCC.
The fourth source of uncertainty is associated with the truncation 
of the expression of $\alpha$ in PRCC theory to second order in cluster 
amplitudes. In one of our earlier studies \cite{mani-10} we observed the 
contribution from third order terms  is negligibly small.
The last, fifth, source of uncertainty is quantum electrodynamical(QED) 
corrections. However in one of our earlier work \cite{chattopadhyay-13b} we 
have shown the contribution from the vacuum polarization correction to the
$\alpha$ is very small and it is less than 0.1\%. 

 For high-$Z$ atoms like Ra, there are two other possible sources of 
theoretical uncertainties. These are: frequency dependent part of the Breit
interaction; and deviation from no-virtual-pair approximation in the atomic
Hamiltonian. To estimate the uncertainty from the first, we do a
series of calculations using GRASP2K \cite{jonsson-13}, which has the option
of including and scaling the frequency dependent part of Breit interaction.
Based on the calculations, for Ra we estimate the upper bound
on the contribution from frequency dependent part of the Breit interaction to 
$\alpha$ as 0.13\%. Similarly, for Ba it is 0.09\% and can be neglected for
the other lighter atoms. To estimate the second possible source--deviation 
from the no-virtual-pair approximation--we consider the leading order diagram 
of $\alpha$ and find that the contribution from diagrams with virtual pairs 
is less than $10^{-4}$ a.u. for Ra. This can be neglected and perhaps, the 
contribution may be higher for the $\alpha$ of highly charged ions.  Combining 
all the sources, we estimate the uncertainty of our results to be 
below $2.9$\%.


\section{Conclusion}
  We have studied the static dipole polarizability of alkaline Earth metal 
atoms. For this we use the PRCC theory, we have developed, to incorporate the 
external perturbation in an atomic or ionic system. We then, examine
the contributions from $T_3^{(0)}$ based on a symmetric representation we 
have developed and defined the corresponding angular momentum representation
suitable for atomic calculations. The RCCSDT theory reported in this work
can be applied to study other properties like hyperfine structure constants,
and extend to one- and two-valence \cite{mani-11} atoms and ions as well. 
These are in progress and we shall report in our future works.

 Our results of the $\alpha$ of alkaline Earth metal atoms
are in very good agreement with the previous theoretical values and it is
also within the uncertainty limit of the experimental data. This is perhaps
not surprising as the PRCC theory has produced excellent results for closed
shell noble gas atoms as well as singly ionized alkali metal ions and 
doubly charged alkaline Earth metal ions 
\cite{chattopadhyay-12a, chattopadhyay-12b, chattopadhyay-13a, 
chattopadhyay-13b}.
We examine the contribution from the Breit interaction for each
of the alkaline Earth metal atoms and find that it is not negligible.
For these atoms, the Breit interaction decreases the value of 
$\alpha$, this is opposite to the trend observed in noble gas atoms.


\begin{acknowledgements}
We thank S. Gautam, Arko Roy and Kuldeep Suthar for useful discussions. The
results presented in the paper are based on the computations using the 3TFLOP
HPC Cluster at Physical Research Laboratory, Ahmedabad.
\end{acknowledgements}


\bibliography{references}{}
\bibliographystyle{apsrev4-1}

\end{document}